\documentclass[aps,prb,twocolumn,longbibliography,superscriptaddress,nobibnotes,showkeys]{revtex4-1}

\usepackage{graphicx}% Include figure files
\usepackage{dcolumn}% Align table columns on decimal point
\usepackage{bm}% bold math
\usepackage[version=3]{mhchem}
\usepackage{color}

\begin{document}

\title{Defect energy levels and persistent luminescence in Cu-doped ZnS}
\author{Khang Hoang}
\email[Corresponding author. E-mail: ]{khang.hoang@ndsu.edu}
\affiliation{Center for Computationally Assisted Science and Technology \& Department of Physics, North Dakota State University, Fargo, ND 58108, United States}
\author{Camille Latouche}
\author{St\'{e}phane Jobic}
\affiliation{Institut des Mat\'{e}riaux Jean Rouxel (IMN), Universit\'{e} de Nantes, CNRS, 2 rue de la Houssini\`{e}re, 44322 Nantes, France}

\date{\today}

\begin{abstract}

Zinc sulfide (ZnS) based materials are widely used in many applications. Yet, due to a lack of detailed knowledge of defect energy levels, the electrical properties and luminescence mechanisms in the materials still give rise to debate. Here, we report a first-principles study of native point defects and impurities in zincblende ZnS using hybrid density-functional calculations. We find that cation and anion vacancies and antisite defects introduce deep defect levels in the band gap and can act as donors or acceptors depending on the position of the Fermi level. The substitutional impurity Cu$_{\rm Zn}$ acts as a deep acceptor and thus does not contribute to p-type conductivity. Substitutional impurities Al$_{\rm Zn}$ and Cl$_{\rm S}$, on the other hand, are shallow donors. More importantly, we identify the isolated (i.e., unassociated) Cu$_{\rm Zn}$ as a source of the green luminescence observed in ZnS-based phosphors and Cu$_{\rm Zn}$--Al$_{\rm Zn}$ and Cu$_{\rm Zn}$--Cl$_{\rm S}$ defect complexes as sources of blue luminescence. The materials may have both green and blue emissions with the relative intensity dependent on the ratio between the unassociated defect and defect complex concentrations, which is also consistent with experimental observations.

\end{abstract}

%\keywords{}

\maketitle


\section{Introduction}\label{sec;intro}

ZnS has been of great interest for various applications, including display technologies, luminescent devices, and solar cells.\cite{Shionoya2006,Xu2018AFM} ZnS phosphors, for example, have a long history, dating back to the discovery of persistent luminescence in ZnS crystals by Th\'{e}odore Sidot in 1866. It is now believed that copper impurities in Sidot's crystals were responsible for the observed phenomenon.\cite{Shionoya2006} ZnS-based phosphors can emit light in the blue, green, and red spectral regions, depending on the presence of certain native point defects and (intentional or otherwise) impurities. Although the materials have been widely used, the structure of defect centers and the associated luminescence mechanisms are still not well understood.

On the theory side, a number of first-principles studies of native defects and impurities--hereafter commonly referred to as defects--in zincblende (cubic) or wurtzite (hexagonal) ZnS have been reported;\cite{Li2012CPL,Varley2013APL,Varley2014JAP,Pham2015PCCP,Faghaninia2016PCCP} however, only limited information on Cu-related defects is available, especially in the case of the zincblende phase. Moreover, some of these studies\cite{Li2012CPL,Pham2015PCCP} are based on density-functional theory (DFT) within the local-density (LDA) or generalized gradient (GGA) approximation\cite{LDA1980,PW91} and/or the DFT+$U$ extension\cite{anisimov1991} where $U$ is the on-site Coulomb correction; these methods are known to have limited predictive power, often due to their inability to reproduce the experimental band gap and thus the position of defect levels in the band gap region. Yet an accurate determination of defect levels in the band gap caused by these defects is crucial to understanding the luminescence of Cu-doped ZnS materials. 

Here, we report a first-principles study of defects and doping in zincblende ZnS using a hybrid DFT/Hartree-Fock approach.\cite{Perdew1996JCP} In addition to native point defects such as cation and anion vacancies, interstitials, and antisites, we consider substitutional Cu, Al, and Cl impurities and their complexes. Copper, aluminum, and clorine are chosen because they are often present in ZnS phosphors as dopants and/or unintentional impurities.\cite{Shionoya2006,Bol2002JL,Corrado2009JPCA,Car2011Nano} The focus of this work is on determining energy levels in the band gap region caused by the defects and identifying defect centers in Cu-doped ZnS phosphors that are responsible for the green and blue luminescence observed in experiments.           

\section{Methods}\label{sec;method} 

Our calculations are based on DFT with the Heyd-Scuseria-Ernzerhof (HSE) hybrid functional,\cite{heyd:8207} as implemented in the Vienna {\it Ab Initio} Simulation Package (\textsc{vasp}).\cite{VASP2} The screening length is set to the default value of 10 {\AA}, whereas the Hartree-Fock mixing parameter is set to 0.32 to match the experimental band gap. The Zn $3d^{10}4s^2$ and S $3s^23p^4$ electrons are treated as valence electrons whereas the inner electrons as core states within the projector augmented wave method.\cite{PAW1} The plane-wave basis-set cutoff is 400 eV and spin polarization is included. In these calculations, the band gap ($E_g$) of zincblende ZnS is 3.66 eV, a direct gap at $\Gamma$, almost identical to the experimental value (3.7 eV \cite{Shionoya2006}). Defects are modeled using a 2$\times$2$\times$2 (64-atom) supercell and a 2$\times$2$\times$2 Monkhorst-Pack $k$-point mesh for the integrations over the Brillouin zone. In the defect calculations, the lattice constant is fixed to the calculated bulk value (5.42 {\AA}) but all the internal coordinates are relaxed (For comparison, the experimental lattice constant is 5.41 {\AA} \cite{Shionoya2006}). Note that spin-orbit coupling (SOC) is not included. We find that the electronic structure of ZnS obtained in HSE is almost identical to that obtained in HSE$+$SOC (see Fig.~S1 in the Supplementary Material); the inclusion of SOC would thus have negligible effects on the energetics of the defects in ZnS. Besides, there is often strong cancellation between different terms in the defect formation energy.

A general native defect, impurity, or defect complex X in charge state $q$ (with respect to the host lattice) is characterized by its formation energy, defined as\cite{Freysoldt2014RMP}
\begin{align}\label{eq:eform}
E^f({\mathrm{X}}^q)&=&E_{\mathrm{tot}}({\mathrm{X}}^q)-E_{\mathrm{tot}}({\mathrm{host}}) -\sum_{i}{n_i\mu_i} \\ %
\nonumber &&+~q(E_{\mathrm{v}}+\mu_{e})+ \Delta^q ,
\end{align}
where $E_{\mathrm{tot}}(\mathrm{X}^{q})$ and $E_{\mathrm{tot}}(\mathrm{host})$ are the total energies of a supercell containing the defect and the defect-free supercell. $\mu_{i}$ is the atomic chemical potential, accounting for the species $i$ either added ($n_i>0$) or removed ($n_i<0$) from the supercell to form the defect and representing the chemical reservoir with which the species is exchanged. $\mu_e$ is the electronic chemical potential, i.e., the Fermi level, which is the energy of the reservoir for electron exchange and, as a convention, referenced to the valence-band maximum (VBM) of the host ($E_{\rm v}$). Finally, $\Delta^q$ is a correction term to align the electrostatic potentials of the bulk and defect supercells and to account for finite-size effects on the total energies of systems with charged defects.\cite{Freysoldt}

The chemical potentials of Zn and S are referenced to the total energy per atom of bulk Zn and S, respectively, and vary over a range determined by the calculated formation enthalpy of ZnS: $\mu_{\rm Zn}+\mu_{\rm S} = \Delta H({\rm  ZnS}) = -1.93$ eV. The extreme Zn-rich and S-rich conditions correspond to $\mu_{\rm Zn} = 0$ eV and $\mu_{\rm S} = 0 $ eV, respectively. The chemical potentials of Cu, Al, and Cl are assumed to be limited by the formation of bulk Cu (or CuS) under the Zn-rich (S-rich) condition, Al$_2$ZnS$_4$, and ZnCl$_2$, respectively. These assumptions result in $\mu_{\rm Cu} = 0$ eV ($-0.47$ eV), $\mu_{\rm Al} = -0.19$ eV ($-3.08$ eV), and $\mu_{\rm Cl} = -3.22$ eV ($-2.25$ eV) under the Zn-rich (S-rich) condition. One, of course, can choose a different set of assumptions which correspond to different experimental conditions. It is noted, however, that our conclusions are not affected by the choice we made here as defect transition levels (see below) are independent of the atomic chemical potentials.

In the calculation of $\Delta^q$, a total static dielectric constant of 8.18 is used. This dielectric constant is calculated as the sum of the electronic contribution (4.67, obtained in calculations using the HSE functional) and the ionic contribution (3.51, obtained with the Perdew, Burke and Ernzerhof version\cite{GGA} of GGA). For comparison, the experimental dielectric constant is reported to be 8.3.\cite{Shionoya2006}

From defect formation energies, one can calculate the {\it thermodynamic} transition level between charge states $q$ and $q'$ of a defect, $\epsilon(q/q')$, defined as the Fermi-level position at which the formation energy of the defect in charge state $q$ is equal to that in charge state $q'$, i.e,\cite{Freysoldt2014RMP}
\begin{equation}\label{eq;tl}
\epsilon(q/q') = \frac{E^f(X^{q}; \mu_e=0)-E^f(X^{q'}; \mu_e=0)}{q' - q},
\end{equation}
where $E^f(X^{q}; \mu_e=0)$ is the formation energy of the defect X in charge state $q$ when the Fermi level is at the VBM ($\mu_e=0$). This $\epsilon(q/q')$ level would be observed in experiments where the defect in the final charge state $q'$ fully relaxes to its equilibrium configuration after the transition. The {\it optical} transition level $E_{\rm opt}^{q/q'}$ is defined similarly but with the total energy of the final state $q'$ calculated using the atomic configuration of the initial state $q$. Clearly, both the thermodynamic and optical transition levels are independent of the choice of the atomic chemical potentials.

\section{Results and Discussion}\label{sec;results}

\begin{figure}%[t]%
\vspace{0.2cm}
\begin{center}
\includegraphics*[width=\linewidth]{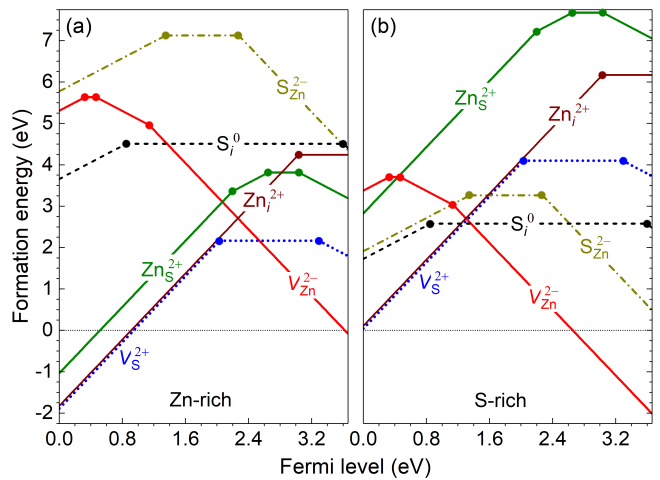}
\end{center}
\vspace{-0.5cm}
\caption{Formation energies of native defects in ZnS under the extreme (a) Zn-rich and (b) S-rich conditions, plotted as a function of the Fermi level from the VBM to the CBM. The slope of energy-line segments indicates the charge state ($q$): positively (negatively) charged defect configurations have positive (negative) slopes; horizontal energy-line segments correspond to neutral defect configurations. Large solid dots connecting two energy-line segments with different slopes mark the {\it thermodynamic} transition levels $\epsilon(q/q')$.}
\label{fig;natives}
\end{figure}

We begin with an examination of native point defects in zincblende ZnS. Figure \ref{fig;natives} shows their calculated formation energies. In general, these defects (except the zinc interstitial, Zn$_i$) are amphoteric, acting as donors or acceptors depending on the position of the Fermi level. The sulfur vacancy, $V_{\rm S}$, is a deep donor with the transition level from the $+2$ to $0$ charge states, $\epsilon(+2/0)$, at 2.03 eV above the VBM. It can also behave as an acceptor under n-type conditions; the $\epsilon(0/-)$ level is at 0.37 eV below the conduction-band minimum (CBM). In the $V_{\rm S}^-$ defect configuration, the electron is delocalized over the four nearest Zn neighbors. The zinc vacancy, $V_{\rm Zn}$, on the other hand, behaves as a deep acceptor; the $\epsilon(-/2-)$ level is at 1.14 eV above the VBM and the $\epsilon(0/-)$ level is at 0.47 eV. The defect can also act as a donor in the range of Fermi-level values closer to the VBM where it can be stable as $V_{\rm Zn}^+$; the $\epsilon(+/0)$ level is at 0.33 eV above the VBM. In the $V_{\rm Zn}^+$ configuration, the electron hole is delocalized over the four nearest S neighbors. Similarly, antisite defects S$_{\rm Zn}$ and Zn$_{\rm S}$ behave as deep donors (acceptors) for Fermi-level values closer to the VBM (CBM). S$_{\rm Zn}$ introduces energy levels at 1.35 eV above the VBM and 1.39 eV below the CBM; Zn$_{\rm S}$ has levels at 0.62, 1.01, and 1.46 eV below the CBM. Regarding the interstitials, S$_i$ introduces two levels in the band gap: $\epsilon(+/0)$ at 0.85 eV above the VBM and $\epsilon(0/2-)$ at 0.05 eV below the CBM. Zn$_i$ is, on the other hand, a deep donor with the $\epsilon(2+/0)$ level at 0.63 eV below the CBM. As discussed later, the energy levels in the band gap region introduced by the native point defects can act as electron or hole traps and likely play a role in the persistent luminescence observed in ZnS-based materials.

\begin{figure}%[t]%
\vspace{0.2cm}
\begin{center}
\includegraphics*[width=\linewidth]{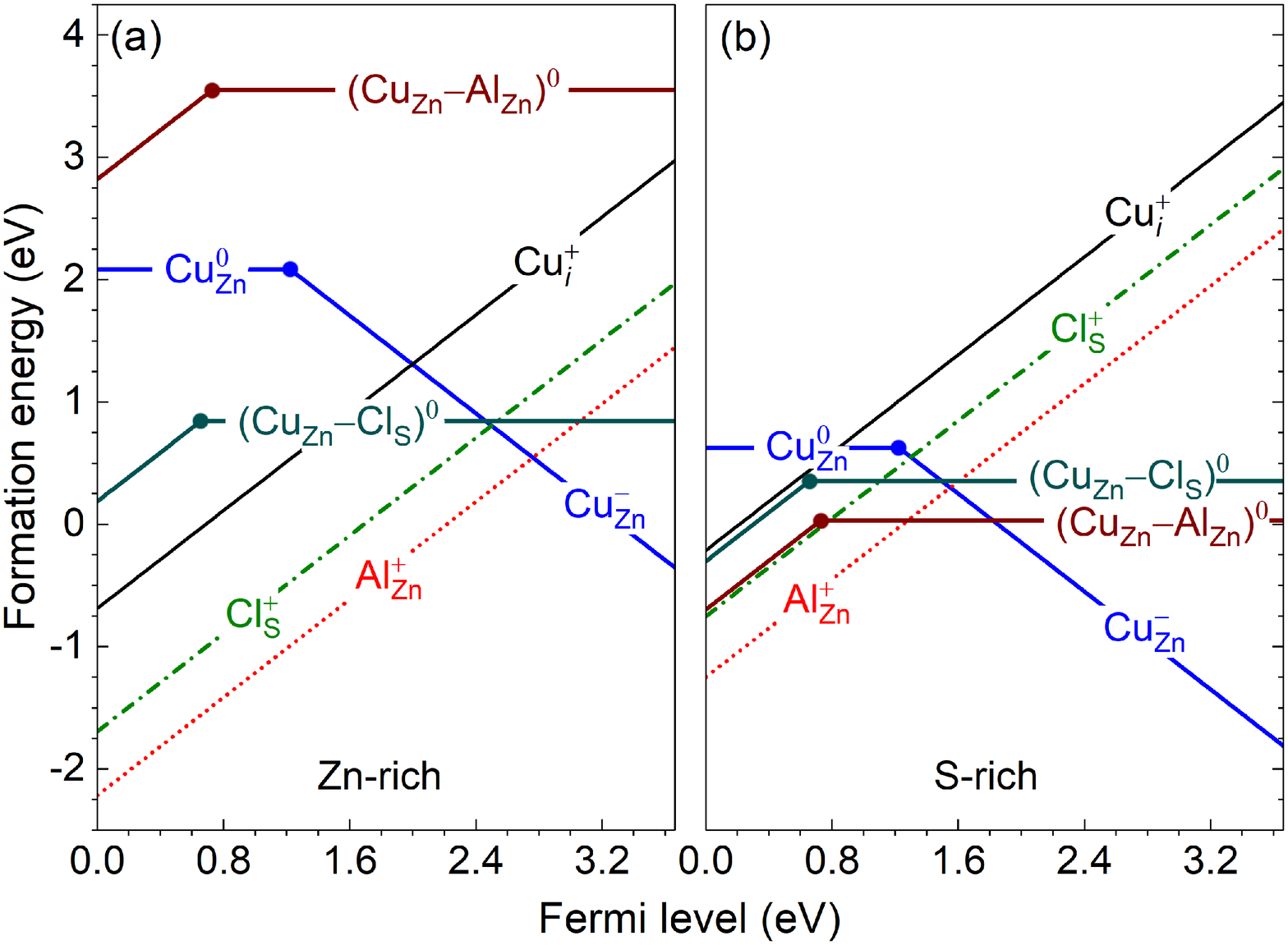}
\end{center}
\vspace{-0.5cm}
\caption{Formation energies of select impurities in ZnS under the extreme (a) Zn-rich and (b) S-rich conditions, plotted as a function of the Fermi level from the VBM to the CBM.}
\label{fig;impurities}
\end{figure}     

Under both Zn-rich and S-rich conditions, the dominant native defects in ZnS are $V_{\rm S}^{2+}$ and Zn$_i^{2+}$ under p-type conditions or $V_{\rm Zn}^{2-}$ under n-type conditions. The formation energies of $V_{\rm S}^{2+}$, Zn$_i^{2+}$, and Zn$_{\rm S}^{2+}$ are even negative in the range of Fermi-level values near the VBM under the Zn-rich condition, indicating that these defects are strong charge-compensating centers. In that context, $V_{\rm S}$ (or Zn$_i$), for example, would ``pin'' the Fermi level at least $\sim$1 eV above the VBM. Under the S-rich condition, the formation energy of $V_{\rm Zn}^{2-}$ is negative for Fermi-level values near the CBM; $V_{\rm Zn}$ would thus ``pin'' the Fermi level at least $\sim$1 eV below the CBM. Under Zn-rich conditions, p-type dopability is thus severely limited by $V_{\rm S}^{2+}$ and Zn$_i^{2+}$; under S-rich conditions, n-type dopability is prohibited by $V_{\rm Zn}^{2-}$. Given the defect landscape of ZnS, n-type (p-type) doping should be carried out under Zn-rich (S-rich) conditions to reduce charge compensation caused by native defects. In general, our results for {\it zincblende} ZnS are quite similar to those for {\it wurtzite} ZnS reported by Varley and Lordi.\cite{Varley2013APL} Differences, where present, can be attributed to the difference in the crystal structure. 

\begin{figure*}%[!hb]%
\vspace{0.2cm}
\includegraphics*[width=\linewidth]{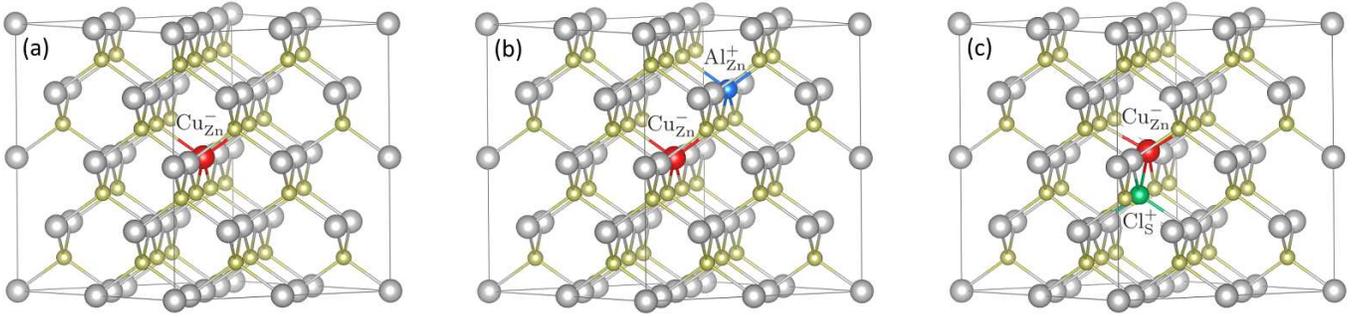}
\caption{Structure of Cu-related defects in zincblende ZnS: (a) Cu$_{\rm Zn}^-$, (b) (Cu$_{\rm Zn}$$-$Al$_{\rm Zn}$)$^0$, i.e., a complex of Cu$_{\rm Zn}^-$ and Al$_{\rm Zn}^+$, and (c) (Cu$_{\rm Zn}$$-$Cl$_{\rm S}$)$^0$, i.e., a complex of Cu$_{\rm Zn}^-$ and Cl$_{\rm S}^+$. Large (gray) spheres are Zn and small (yellow) spheres are S.}
\label{fig;struct}
\end{figure*}

Next, we examine the effects of impurities, intentionally incorporated or otherwise. Figure \ref{fig;impurities} shows the calculated formation energies of ZnS doped with Cu, Al, or Cl, or co-doped with (Cu,Al) or (Cu,Cl). The substitutional Cu impurity, Cu$_{\rm Zn}$, is stable as Cu$_{\rm Zn}^0$ (i.e., Cu$^{2+}$) under p-type conditions or Cu$_{\rm Zn}^-$ (i.e., Cu$^+$) under n-type conditions and introduces a deep acceptor $\epsilon(0/-)$ level at 1.22 eV above the VBM. This calculated value is in excellent agreement with the experimental one (1.25 eV) reported in the literature.\cite{Shionoya2006} The result also indicates that Cu$_{\rm Zn}$ does not contribute to p-type conductivity, at least at not too high doping levels. The atomic structure of Cu$_{\rm Zn}^-$ is presented in Fig.~\ref{fig;struct}(a). The impurity is tetrahedrally coordinated with sulfur; the Cu--S bond length is 2.333 {\AA} ($\times 4$), compared to 2.345 {\AA} of the Zn--S bonds in bulk ZnS. 

The Cu interstitial, Cu$_i$, is stable as Cu$_i^+$ (i.e., Cu$^+$) in the entire range of Fermi-level values. In Cu$_i^+$, the Cu$^+$ ion forms chemical bonds with four S atoms with the bond length of 2.360--2.366 {\AA}. The defect is thus a shallow donor and can, in principle, cause n-type conductivity in the material. Its concentration under n-type conditions is, however, expected to be negligible compared to that of Cu$_{\rm Zn}^-$ due to the large difference in the formation energies; see Fig.~\ref{fig;impurities}. This appears to be consistent with experimental reports showing that there are negligible or no Cu interstitials in Cu-doped ZnS nanocrystals.\cite{Corrado2009JPCA,Car2011Nano}   

Regarding Al$_{\rm Zn}$ and Cl$_{\rm S}$, they are stable as Al$_{\rm Zn}^+$ and Cl$_{\rm S}^+$, respectively, over the entire range of Fermi-level values. Like in the case of Cu$_i$, these defects donate one electron to the host and become ionized; the electron then transfers to the CBM. Al$_{\rm Zn}$ and Cl$_{\rm S}$ are thus shallow donors, which is consistent with experimental observations.\cite{Yamaga1988JCG,Yasuda1997JCG} In Al$_{\rm Zn}^+$, the Al--S bond length is 2.272 {\AA} ($\times 4$), whereas in Cl$_{\rm S}^+$ the local lattice environment is slightly more distorted with Cl--Zn bonds of 2.559 {\AA} ($\times 3$) and 2.563 {\AA}. These impurities are often found in ZnS-based luminescent materials.\cite{Shionoya2006,Corrado2009JPCA} Note that the n-type conductivity\cite{Shionoya2006} usually observed in as-grown single crystals of ZnS could be due to the presence of unintentional Al$_{\rm Zn}$ or Cl$_{\rm S}$ impurities. The observed conductivity may also be caused by the presence of other (intentional or otherwise) impurities not yet considered here. For example, it has been reported that hydrogen incorporated on the S site, H$_{\rm S}$, behaves as a shallow donor in {\it wurtzite} ZnS.\cite{Varley2013APL}    

The above impurities, especially those with oppositely charged defect configurations, can form complexes. Explicit calculations are therefore carried out for Cu$_{\rm Zn}$--Al$_{\rm Zn}$ and Cu$_{\rm Zn}$--Cl$_{\rm S}$. We find that these complexes are stable in the neutral charge state under n-type conditions or in the $+$ state under p-type conditions. The $\epsilon(+/0)$ level is at 0.73 eV above the VBM for Cu$_{\rm Zn}$--Al$_{\rm Zn}$ and 0.66 eV for Cu$_{\rm Zn}$--Cl$_{\rm S}$. The defect level is thus shifted, compared to that in the isolated Cu$_{\rm Zn}$, indicating strong local elastic and electrostatic interactions between the constituents in the complexes. In general, the electronic behavior of defect complexes can be significantly different from that of their isolated defects, a phenomenon also observed in other materials.\cite{Yan2012APL,Hoang2016RRL}

Figures \ref{fig;struct}(b) and \ref{fig;struct}(c) show the structure of (Cu$_{\rm Zn}$--Al$_{\rm Zn}$)$^0$ and (Cu$_{\rm Zn}$--Cl$_{\rm S}$)$^0$, respectively. In (Cu$_{\rm Zn}$--Al$_{\rm Zn}$)$^0$, the Cu--S bond lengths are 2.322 {\AA} ($\times 2$), 2.324 {\AA}, and 2.376 {\AA}; the Al--S bond lengths are 2.281 {\AA ($\times 2$), 2.240 {\AA}, and 2.284 {\AA}. The binding energy is 0.57 eV with respect to the isolated (i.e., unassociated) constituents Cu$_{\rm Zn}^-$ and Al$_{\rm Zn}^+$ in the Fermi-level range $\mu_e>0.73$ eV (and hence under n-type conditions). In (Cu$_{\rm Zn}$--Cl$_{\rm S}$)$^0$, the Cu--S bond length is 2.258 {\AA} ($\times 3$) and the Cl--Zn bond length is 2.471 {\AA ($\times 3$); the Cu--Cl distance is 2.876 {\AA}. The binding energy is 0.77 eV with respect to the isolated Cu$_{\rm Zn}^-$ and Cl$_{\rm S}^+$ in the Fermi-level range $\mu_e>0.66$ eV (and hence under n-type conditions). In both cases, the defect centers thus have {\it lower symmetry} than the host lattice. The binding energies are small, which suggests that the concentration of the complexes is likely much smaller than that of their isolated constituents when the material is prepared under thermodynamic equilibrium. The defect complexes are even less stable under p-type conditions: the binding energy of (Cu$_{\rm Zn}$--Al$_{\rm Zn}$)$^+$ is 0.08 eV with respect to isolated Cu$_{\rm Zn}^0$ and Al$_{\rm Zn}^+$ (in the range $\mu_e<0.73$ eV) and that of (Cu$_{\rm Zn}$--Cl$_{\rm S}$)$^+$ is 0.21 eV with respect to isolated Cu$_{\rm Zn}^0$ and Cl$_{\rm S}^+$ ($\mu_e<0.66$ eV).

Given the calculated defect levels, the Cu-related defects can play a role in high-energy luminescent transitions in ZnS. An electron previously excited from the valence band to the conduction band or one at the shallow donor level such as that associated with Al$_{\rm Zn}^+$ or Cl$_{\rm S}^+$ in n-type ZnS, for example, can recombine with the empty defect state of the isolated Cu$_{\rm Zn}^0$, resulting in a Cu$_{\rm Zn}^0$ $\rightarrow$ Cu$_{\rm Zn}^-$ transition. The corresponding peak emission energy is $E_g - E_{\rm opt}^{0/-} = 2.28$ eV, which is in the green region of the spectrum, with a relaxation energy of 0.16 eV; here, $E_{\rm opt}^{0/-}$ is given by the formation-energy difference between Cu$_{\rm Zn}^0$ and Cu$_{\rm Zn}^-$, both in the lattice configuration of the initial configuration; see also Sec.~\ref{sec;method}. For the Cu$_{\rm Zn}$--Al$_{\rm Zn}$ and Cu$_{\rm Zn}$--Cl$_{\rm S}$ complexes, the peak emission energies associated with the transition from the $+$ state to the neutral state is $E_g - E_{\rm opt}^{+/0} =$ 2.79 eV and 2.81 eV with relaxation energies of 0.14 eV and 0.20 eV, respectively. Both the energies are thus in the blue region of the spectrum. 

The isolated Cu$_{\rm Zn}$ (i.e., Cu$_{\rm Zn}$ that is not bound to a co-activator such as Al$_{\rm Zn}$ or Cl$_{\rm S}$) is therefore a source of green luminescence in Cu-doped ZnS phosphors and Cu$_{\rm Zn}$--Al$_{\rm Zn}$ and Cu$_{\rm Zn}$--Cl$_{\rm S}$ complexes are sources of blue luminescence. As indicated above, the emission in all these cases is assumed to take place from the CBM to the (optical) transition level of the defect centers. The hydrogenic effective-mass state associated with a shallow donor such as Al$_{\rm Zn}$ or Cl$_{\rm S}$ is expected to be very close to the CBM. A schematic representation of these optical transitions and other energies, including the zero-phonon line (ZPL), is given in Fig.~S2 in the Supplementary Material. Experimentally, it has been observed that the blue luminescent centers have lower symmetry than the host lattice whereas the symmetry is not lower in the case of the green luminescent center,\cite{Shionoya2006,Corrado2009JPCA} which is consistent with our results and the analysis of the local lattice environment of Cu$_{\rm Zn}$ vs.~Cu$_{\rm Zn}$--Al$_{\rm Zn}$ and Cu$_{\rm Zn}$--Cl$_{\rm S}$ presented earlier.     

The persistence of the luminescence observed in ZnS-based phosphors\cite{Shionoya2006} could be due to electron trapping mediated by native defects. An electron excited from the valence band to the conduction band, for example, can be trapped in one of the defect levels $V_{\rm S}$, Zn$_i$, or Zn$_{\rm S}$ introduce below the CBM (see Fig.~\ref{fig;natives}). The electron can then be released back to the conduction band through some detrapping mechanism (e.g., thermal activation) and participate in the optical transitions discussed above. Whether $V_{\rm S}$, Zn$_i$, and/or Zn$_{\rm S}$ are present in ZnS samples with significant concentrations likely depend on the synthesis conditions. Note that defects with a high formation energy may still form with high concentrations when the material is prepared under highly non-equilibrium conditions. In this case, a defect's concentration is not directly determined by its formation energy. 

Finally, in the (Cu,Al) and (Cu,Cl) doping, it is likely that both the isolated defects and defect complexes are present. As a result, the ZnS-based materials can exhibit both green and blue luminescence with different intensities. The concentration ratio between the unassociated Cu$_{\rm Zn}$ and the Cu$_{\rm Zn}$--Al$_{\rm Zn}$ or Cu$_{\rm Zn}$--Cl$_{\rm S}$ complex is expected to depend on the Cu/Al or Cu/Cl ratio in the environment as well as other synthesis conditions. Experimentally, Chen et al.,~\cite{Chen2001TSF} for example, found that the green (blue) emission is dominant at low (high) Cu concentrations. The observation can be understood as the following: At low Cu doping levels, the probability of Cu being close to the co-dopant (Al or Cl) is low; i.e., Cu is present in the material predominantly as the isolated Cu$_{\rm Zn}$. The probability of Cu being close to Al or Cl is, however, high at high Cu doping levels; as a result, Cu can be present predominantly in the defect complex form, Cu$_{\rm Zn}$--Al$_{\rm Zn}$ or Cu$_{\rm Zn}$--Cl$_{\rm S}$. It is the defect--defect interaction within the defect complex that causes the shift from green to blue in the luminescence of these doped ZnS materials.

\section{Conclusions} 

We have investigated native point defects and impurities in zincblende ZnS using hybrid density-functional calculations. Cation and anion vacancies and antisite defects are found to introduce deep defect levels in the band gap region and act as donors or acceptors depending on the position of the Fermi level. The Al and Cl impurities act as shallow donors whereas the Cu-related defects introduce deep defect levels. Significantly, we identify the unassociated Cu$_{\rm Zn}$ is a source of green luminescence in ZnS-type phosphors and Cu$_{\rm Zn}$--Al$_{\rm Zn}$ and Cu$_{\rm Zn}$--Cl$_{\rm S}$ defect complexes are sources of blue luminescence.

\begin{acknowledgments}

K.H.~is grateful to Universit\'{e} de Nantes for supporting his visit to Institut des Mat\'{e}riaux Jean Rouxel (IMN) during which this work was initiated. This work used resources of the Extreme Science and Engineering Discovery Environment (XSEDE), which is supported by National Science Foundation Grant No.~ACI-1548562, the National Energy Research Scientific Computing Center (NERSC), a U.S.~Department of Energy Office of Science User Facility operated under Contract No.~DE-AC02-05CH11231, the Center for Computationally Assisted Science and Technology (CCAST) at North Dakota State University, and Centre de Calcul Intensif des Pays de Loire (CCIPL), France.

\end{acknowledgments}

%\bibliography{optoelectronics}
%

\end{document}